\documentclass[aps,twocolumn]{revtex4}

\usepackage{graphicx}
\begin{document}

\title{Modeling spin transport with current-sensing spin detectors}

\author{Jing Li}
\affiliation{Center for Nanophysics and Advanced Materials, Department of Physics, University of Maryland, College Park MD 20742 USA}
\author{Ian Appelbaum}
\altaffiliation{appelbaum@physics.umd.edu}
\affiliation{Center for Nanophysics and Advanced Materials, Department of Physics, University of Maryland, College Park MD 20742 USA}

\begin{abstract}
By incorporating the proper boundary conditions, we analytically derive the impulse response (or ``Green's function'') of a current-sensing spin detector. We also compare this result to a Monte-Carlo simulation (which automatically takes the proper boundary condition into account) and an empirical spin transit time distribution obtained from experimental spin precession measurements. In the strong drift-dominated transport regime, this spin \emph{current} impulse response can be approximated by multiplying the spin \emph{density} impulse response by the average drift velocity. However, in weak drift fields, large modeling errors up to a factor of 3 in most-probable spin transit time can be incurred unless the full spin current Green's function is used.  
\end{abstract}

\maketitle

Measuring spin-polarized electron transport requires both successful spin injection and spin detection.\cite{ZUTICRMP, APS, ZUTICNATURE} To extract meaningful spin transport parameters, modeling the response characteristics of these components is essential. In particular, the details of the spin detection mechanism can impose very different boundary conditions on the spin distribution and affect the device output response. 

Two types of electrical spin-detection techniques are currently in use: voltage sensing\cite{JOHNSON, JEDEMA, LOU, JONKERLATERAL, VANWEESGRAPHENE} and current sensing\cite{APPELBAUMNATURE, BIQINPRL, 35PERCENT, BIQINJAP, 2MM, SPINFETEXPT, DEPHASING}. Because voltage sensing (which is sensitive to the spin \emph{density}) uses an open-circuit configuration, ideally it does not serve as a sink for spins and therefore the presence of the detector does not impose any boundary conditions on the spin distribution. However, current sensing methods (which are sensitive to spin \emph{current} flowing into the contact) are ideally perfect sinks of charge and spin, and so do indeed impose a boundary condition. Specifically, they should constrain the spin density to vanish at the detector contact, so the functional response of the two detectors will be very different. Here, we show how this boundary condition can be incorporated into the problem to accurately model current-sensing spin detectors, and find the regime under which large simulation errors are incurred by ignoring it.
  
To model the spin detector, the transport impulse response (or ``Green's function'', i.e. spatio-temporal evolution of spin density with initial conditions $s(x,t=0)=\delta (x)$, where $\delta (x)$ is the Dirac-delta function) is required. The spin density $s(x,t)$ for $0<x<L$ (where $L$ is spin transit length; the injector is at $x=0$ and the detector at $x=L$) and $t>0$ is determined by the spin drift-diffusion equation 

\begin{equation} 
\frac{\partial s}{\partial t}=D\frac{\partial^2 s}{\partial x^2}-v\frac{\partial s}{\partial x}-\frac{s}{\tau},
\label{DRIFTDIFFEQN}
\end{equation} 

\noindent where $v$ is the drift velocity, $D$ is the diffusion coefficient, and $\tau$ is the spin lifetime. Once we calculate this Green's function, we can construct the response to arbitrary spin injection conditions, even the DC currents used so far in experiments.

With voltage-sensing spin detection, there are no boundary conditions on $s(x,t)$ (besides those at infinity which keep the solution bounded). Assuming $v$ is a constant independent of position $x$, the Green's function of Eqn. \ref{DRIFTDIFFEQN} can be determined straightforwardly using e.g. separation of variables and the Fourier representation of the Dirac delta function. It is given by

\begin{equation}
s(x,t)=\frac{1}{2\sqrt{\pi Dt}} e^{-\frac{(x-vt)^2}{4Dt}}e^{-t/\tau}.
\label{NOBCSGREENSEQ}
\end{equation}  

\begin{figure}
\includegraphics[width=6cm,height=6cm]{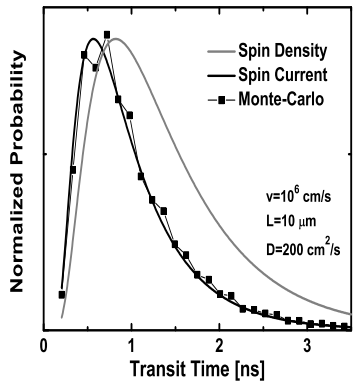}
\caption{\label{DISTRFIG} 
Comparison of the time dependence of spin density at the detector using a voltage-sensing contact, spin current using a current sensing contact (which imposes a boundary condition on spin density), and a direct Monte-Carlo distribution which automatically imposes the same boundary condition. Parameters $v=10^6$cm/s, $D=200$cm$^2$/s and $L=10\mu$m have been used to emphasize the relative differences, which are proportional to $2D/Lv$ to first-order and therefore decrease rapidly for higher drift velocity and longer transit length.}
\end{figure}

\noindent The impulse response at the detector can be obtained simply by substituting $x=L$. An example spin density distribution using Eqn. \ref{NOBCSGREENSEQ} and $v=10^6$cm/s, $L=10\mu$m, and $D=200$cm$^2$/s is shown in Fig. \ref{DISTRFIG}, where the effects of spin relaxation with finite $\tau$ are ignored.

However, the Green's function of Eqn. \ref{DRIFTDIFFEQN} in Eqn. \ref{NOBCSGREENSEQ} is not consistent with current-sensing detection. Unlike in open-circuit voltage spin detection, electrons cross into the detector contact. Once they cross the transport channel boundary, they do not return via diffusion, especially in ballistic hot-electron spin detectors where inelastic scattering prevents it\cite{APPELBAUMNATURE, BIQINPRL, 35PERCENT, BIQINJAP, 2MM, SPINFETEXPT, DEPHASING}. This imposes an absorbing boundary condition on the spin density at the detector ($s(x=L,t)=0$), from which spin current is determined by Fick's law, 

\begin{equation}
J_s(x=L,t)=-D\frac{ds}{dx}|_{x=L}.
\label{FICK}
\end{equation}

The Green's function satisfying this boundary condition can be found by making an ansatz equivalent to the method of images so that we can construct the solution using Eqn. \ref{NOBCSGREENSEQ}. The \emph{negative-valued} ``image'' spin density also moves at drift velocity $v$ but its relative magnitude is determined by the transport parameters such that the boundary condition $s(x=L,t)=0$ is satisfied:

\begin{equation}
s(x,t)=\frac{1}{2\sqrt{\pi Dt}} \left[ e^{-\frac{(x-vt)^2}{4Dt}}- e^{Lv/D} e^{-\frac{(x-2L-vt)^2}{4Dt}} \right]e^{-t/\tau}.
\label{GREENSEQ}
\end{equation} 

\noindent The corresponding spin current at the detector, obtained using Eqn. \ref{FICK}, is then

\begin{equation}
J_s(x=L,t)=\frac{1}{2\sqrt{\pi Dt}}\frac{L}{t}e^{-\frac{(L-vt)^2}{4Dt}}e^{-t/\tau}.
\label{SPINCURREQ}
\end{equation}

\begin{figure}
\includegraphics[width=6cm,height=6cm]{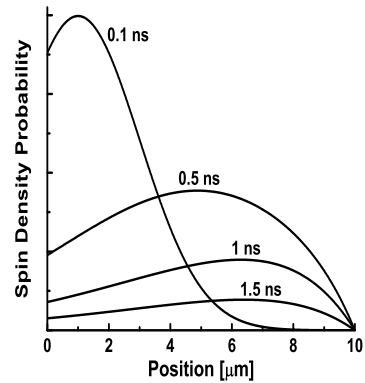}
\caption{\label{GREENSFIG} Evolution of the Green's function of the spin drift-diffusion equation, where a Dirac delta spin density distribution is injected at $x=0$ at $t=0$, for time increments shown in the legend. Drift velocity is $10^6$cm/s and diffusion coefficient is 200cm$^2$/s. For clarity, no spin depolarization over this timescale is considered. Note the effect of absorbing boundary conditions at the detector ($x=10\mu$m). The corresponding spin current distribution, given by Eqn. \ref{SPINCURREQ}, is shown in Fig. \ref{DISTRFIG}.
}
\end{figure}

\noindent The time-evolution of Eqn. \ref{GREENSEQ} (using the same values for $v$, $D$, and $L$ as in Fig. \ref{DISTRFIG}) is shown in Fig. \ref{GREENSFIG}. The corresponding spin current, obtained by using Eqn. \ref{FICK} and given by Eqn. \ref{SPINCURREQ}, is compared to the spin density from Eqn. \ref{NOBCSGREENSEQ} in Fig. \ref{DISTRFIG}. Significant differences between the qualitative form and position of the two functions are evident with these particular transport parameters.

Note that the expression in Eqn. \ref{SPINCURREQ} is simply the spin density in the absence of the boundary condition (Eqn. \ref{NOBCSGREENSEQ}), multiplied by the spin velocity $L/t$. This is a particularly satisfying result in light of the normal definition of particle current. 

Because current-sensing measurements are typically under conditions of strong drift fields (where $v>>D/L$)\cite{APPELBAUMNATURE, BIQINPRL, 35PERCENT, BIQINJAP, 2MM, SPINFETEXPT, DEPHASING}, we have approximated this $L/t$ term with a constant (e.g. the average drift velocity $v$) in previous work comparing experiment to theory.\cite{SPINFETTHRY, DEPHASING, BIQINPRL} This approximation is justified in the strongly drift-dominated regime, because it involves only a rigid shift of the most probable transit time from $\approx L/v-D/v^2+2D^2/v^3L$ to $\approx L/v-3D/v^2+9D^2/2v^3L$, a difference of $2D/v^2$ to first order and a subsequent error in the measured velocity of $2D/L$. Because typical values for these variables are on the order of $v >10^6$ cm/s, $L>10^{-2}$cm, and $10^2<D<10^3$cm$^2$/s, the relative error associated with this approximation ($2D/Lv$ in both cases) is then just a few percent. 

However, the error can be substantial at low drift velocity when diffusion is the dominant transport mechanism. In the case that $D>>Lv$, use of the density Green's function (Eqn. \ref{NOBCSGREENSEQ}), without incorporation of the functional dependence of the $L/t$ term, will give a calculated most-probable (peak) transit time of $L^2/2D$, whereas the correct spin current distribution (Eqn. \ref{SPINCURREQ}) will give a value three times smaller. This large potential error highlights the importance of using the appropriate Green's function to model the spin transport, even when weak drift fields are used. In addition, it should be noted here that in diffusion-dominant transport, the current-sensing detector sees a response which has a smaller standard deviation $\Delta t$, resulting in smaller spin dephasing and therefore higher spin coherence than voltage-sensing detectors.

A consistency check of this spin-current impulse response derived above can be provided by direct modeling of the drift and diffusion of an ensemble of electrons. This simulation can be used to assemble a histogram of transit times, which in the limit of large numbers of electrons yields the transit-time distribution function. Such a ``Monte Carlo'' method, which incorporates the proper boundary conditions for our current-based spin detection technique automatically (because electron transport simulation ceases once $x>L$), was already used to model spin transport through doped Si\cite{DOPED}, where Eqn. \ref{DRIFTDIFFEQN} no longer has constant coefficents due to an $x$-dependent drift velocity caused by inhomogeneous electric fields from ionized dopant impurities. 

The Monte-Carlo technique involves discretely stepping through time a duration $\delta t$, modeling drift with a spatial translation $v\delta t$, and diffusion with a translation of $\pm\sqrt{2D\delta t}$ (where the sign is randomly chosen to model the stochastic nature of the process), until $x>L$. Then, the total transit time is recorded. This process is repeated many times until the list of transit times can be used to make a reliable histogram.

A simulated Monte-Carlo distribution using $4\times 10^4$ electron drift-diffusion trajectories and $\delta t=10^{-11}$s is compared to the analytic functions of spin density and spin current (using the same transport parameters $v$, $D$, and $L$) in Fig. \ref{DISTRFIG}. Clearly, the Monte-Carlo distribution closely resembles the spin current distribution, confirming that this method does indeed automatically incorporate the appropriate absorbing boundary conditions.

\begin{figure}
\includegraphics[width=7.5cm,height=7cm]{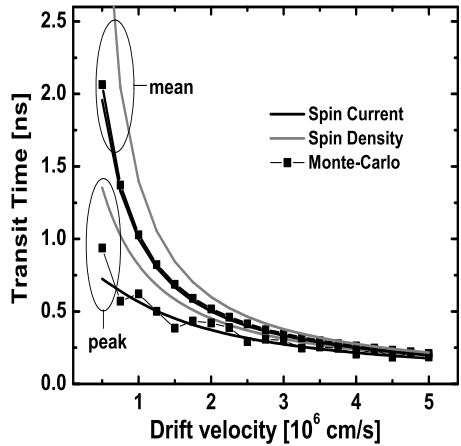}
\caption{\label{FIG3} 
Comparison of most-probable (peak) and average (mean) spin transit times calculated using spin density without boundary conditions (Eqn. \ref{NOBCSGREENSEQ}), spin current with absorbing boundary conditions (Eqn. \ref{SPINCURREQ}), and using a Monte-Carlo method. Transport parameters are the same as those used in Figs. 1 and 2.
}
\end{figure}

Fig. \ref{FIG3} shows the drift velocity dependences of the most probable (peak) and average (mean) transit times of the spin density, spin current, and Monte-Carlo distributions using the same transport parameters as in Fig. \ref{DISTRFIG}. Clearly, the differences between the correct (spin current and Monte-Carlo) impulse response distributions and the spin-density distribution becomes rapidly smaller as $v$ increases. It should also be noted that for transit lengths $L$ larger than the 10$\mu$m used here, the differences are likewise smaller due to the functional form of relative error ($D/Lv$).

\begin{figure}[h!]
\includegraphics[width=7.5cm,height=7cm]{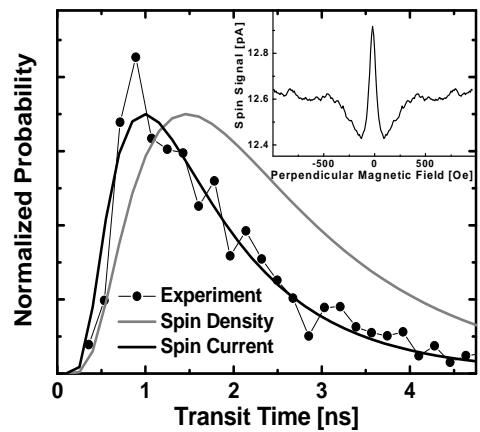}
\caption{\label{FIG4} 
Comparison of empirical spin current transit time distribution with spin density calculated without boundary conditions (Eqn. \ref{NOBCSGREENSEQ}) and spin current with absorbing boundary conditions (Eqn. \ref{SPINCURREQ}). Inset: Spin precession data used to calculate empirical distribution with Fourier transform.\cite{LATERAL} Shown is the average of four perpendicular magnetic field sweeps, symmetrized to avoid ferromagnetic contact magnetization switching and constrain the transform to be real. 
}
\end{figure}

As has been shown elsewhere, the real part of the Fourier transform of spin precession data measured in a perpendicular magnetic field reveals the empirical transit time distribution.\cite{LATERAL} We can therefore compare experimentally-obtained distributions to the analytically-derived models here with data from 10$\mu$m-thick undoped silicon spin transport devices utilizing current-sensing spin detectors\cite{APPELBAUMNATURE, 35PERCENT}, as in Fig. \ref{FIG4}. Experimental results shown in the inset are obtained with an internal electric field of approximately 50V/cm at a temperature of 90K. Because of a rapidly degrading signal-to-noise, this is nearly the lowest electric field and hence the weakest drift conditions that can be reliably measured with these devices. Fitting parameters used to produce the analytic spin current (Eqn. \ref{SPINCURREQ}) and spin density (Eqn. \ref{NOBCSGREENSEQ}) are $v=5.5\times 10^6$cm/s and $D=$120 cm$^2$/s (consistent with $D$ for measurements at higher fields). The empirical distribution closely matches the spin current distribution, again confirming the importance of using Eqn. \ref{SPINCURREQ} in modeling current-sensing spin detectors.

In summary, we have derived the correct drift-diffusion Green's function for modeling the response of current-sensing spin detectors and favorably compared it to Monte-Carlo and empirical spin transit time distributions. In the strongly drift-dominated regime, the correct Green's function can be approximated by ignoring the boundary condition at the spin detector and simply multiplying the spin density Green's function by the average drift velocity $v$. However, in weak drift fields or in conditions of strong diffusion, large modeling errors can be made by using the wrong distribution.

We thank B. van Wees and T. Stanescu for important discussions on this topic. 

This work has been supported by the Office of Naval Research and the National Science Foundation.

\end{document}